\documentstyle[graphicx]{jaa}
\def\kms {\ifmmode{{\rm ~km~s}^{-1}}\else{~km~s$^{-1}$}\fi}

\def\lsun {\ifmmode{{\rm ~L}_\odot}\else{~L$_\odot$}\fi}
\def\deg {^{\circ}}
\def\sqdeg {\,deg$^2$}
\def\arcsec {\,arcsec}

\def\ujybm {\,$\mu$Jy/bm}

\def\apj {{\it Ap.~J.\ }}
\def\apjl {{\it Ap.~J.\ (Letters)}}
\def\apjs {{\it Ap.~J.\ Suppl.}}
\def\aj {{\it A.~J.}}
\def\aa {{\it Astr.~Ap.}}
\def\aap {{\it Astr.~Ap. }}

\def\mnras {{\it MNRAS}}

\def\pasa {{\it PASA}}

\def\procspie {{\it Proc. SPIE}}

\def\ngal {68 }

\def\nsf {51 }

\begin{document}
\title{Evolutionary Map of the Universe: Tracing Clusters to High Redshift}
\author[Ray P. Norris and the EMU team]%
       {Ray P. Norris\thanks{e-mail:Ray.Norris@csiro.au} 
       \& the EMU team\thanks{listed on http://askap.pbworks.com/TeamMembers}\\
        CSIRO Astronomy \& Space Science, PO Box 76, Epping, NSW, Australia}

\maketitle
\label{firstpage}
\begin{abstract}
The Australian SKA Pathfinder (ASKAP) is a new radio-telescope being built in Western Australia. One of the key surveys for which it is being built is EMU (Evolutionary Map of the Universe), which will make a deep ($\sim$10 \ujybm\ rms) radio continuum survey covering the entire sky as far North as $+30\deg$. EMU may be compared to the NRAO VLA Sky Survey (NVSS), except that it will have about 45 times the sensitivity, and five times the resolution.  EMU will also have much better sensitivity to diffuse emission than previous large surveys, and is expected to produce a large catalogue of relics, tailed galaxies, and haloes, and will increase the number of known clusters by a significant factor. Here we describe the EMU project and its impact on the astrophysics of clusters.
\end{abstract}


\section{Introduction}
\label{sec:intro}
The Australian SKA Pathfinder (ASKAP: Johnston et al., 2007,2008; DeBoer et al., 2009) is a new radio telescope 
under construction in Western Australia. Not only will ASKAP be a technology  pathfinder for the Square Kilometre Array (SKA), but it will also be a major survey telescope in its own right, likely to generate significant new astronomical discoveries, through projects such as EMU (Evolutionary Map of the Universe). 
In this paper, we describe ASKAP in \S 2, then EMU in \S 3. \S 4 discusses one of the science goals of EMU, which is to detect and study clusters of galaxies.
Full details of EMU, including a discussion of all the science goals, the techniques being developed to achieve them, and the plans for the EMU survey, can be found in Norris et al. (2011). 
\section{ASKAP}
ASKAP comprises 36 12-metre antennas spread over a region 6-km in diameter, each equipped with a novel phased-array feed (PAF) of 96 dual-polarisation pixels, operating  in the 700--1800 MHz band, giving  ASKAP  a  field of view (FOV) of $\sim$ 30 \sqdeg.
The ASKAP array configuration  (Gupta, 2008) includes a central core of 30 antennas distributed over a region $\sim$\,700 m in diameter, corresponding to a point spread function of $\sim$ 30\arcsec~using natural weighting, and a further six antennas arranged with a maximum baseline of 6 km, corresponding to a point spread function of $\sim$ 10\arcsec~ using uniform weighting. The array gives excellent \emph{uv} coverage between declination $-90\deg$ and $+30\deg$.

The outputs of the 96 dual-polarisation receivers are combined in a beam-former to form up to 36 beams within a 30-\sqdeg\ envelope.  The EMU observing strategy will be to observe one field for 12 hours, reaching an rms sensitivity of $\sim$ 10\ujybm\  over a $\sim$ 30\sqdeg\ FOV, with a uniformity (after dithering) of $\sim$2\% , so that the images from the 36 beams can be imaged and deconvolved as a single image covering the FOV.
Because of the short spacings of ASKAP, the $\sim$ 10\ujybm\ rms continuum sensitivity in 12 hours is approximately constant for beam sizes from 10 to 30 arcsec, then increases to $\sim$ 20\ujybm\ for a 90 arcsec beam and $\sim$ 40\ujybm\ for a 3 arcmin beam.

The high ASKAP data rate ($\sim$ 2.5 GB/s) requires processing (including calibration, imaging, and source-finding: Cornwell et al., 2011) in an automated pipeline processor. Initial observations will produce a global sky model (an accurate description of all sources stronger than $\sim$ 1 mJy) which will then be subtracted from the visibility data before processing. This sky model also enables self-calibration of all fields  without any need for separate calibration observations. EMU will observe a 300 MHz band, split into 1 MHz channels, with full Stokes parameters measured in each channel. As well as producing images and source catalogues, the processing pipeline will measure spectral index, spectral curvature, and all polarisation products across the band.
All ASKAP data will be placed in the public domain after quality control, and should be available to the astronomical community from the end of 2013 onwards.

\section{EMU}
Two projects to survey the entire visible sky, EMU and WALLABY, will dominate the initial usage of ASKAP, and are primarily driving its design, with a further eight projects (listed on http://askap.org) also being supported.  EMU (Evolutionary Map of the Universe: Norris et al., 2011) is a 20-cm continuum survey whilst WALLABY (Koribalski et al., 2011) is a survey for neutral hydrogen.   EMU and WALLABY are expected to observe commensally, i.e., they will observe the sky in both continuum and HI modes at the same time, splitting the two data streams into separate processing pipelines. 
 
 Fig. 1 shows how EMU compares with other major 20-cm continuum radio surveys.  All current surveys are bounded by a diagonal line that roughly marks the limit of available telescope time of current-generation radio telescopes. The region to the left of this line is currently unexplored.
 
\begin{figure}[h]
\begin{center}
\includegraphics[scale=0.4, angle=0]{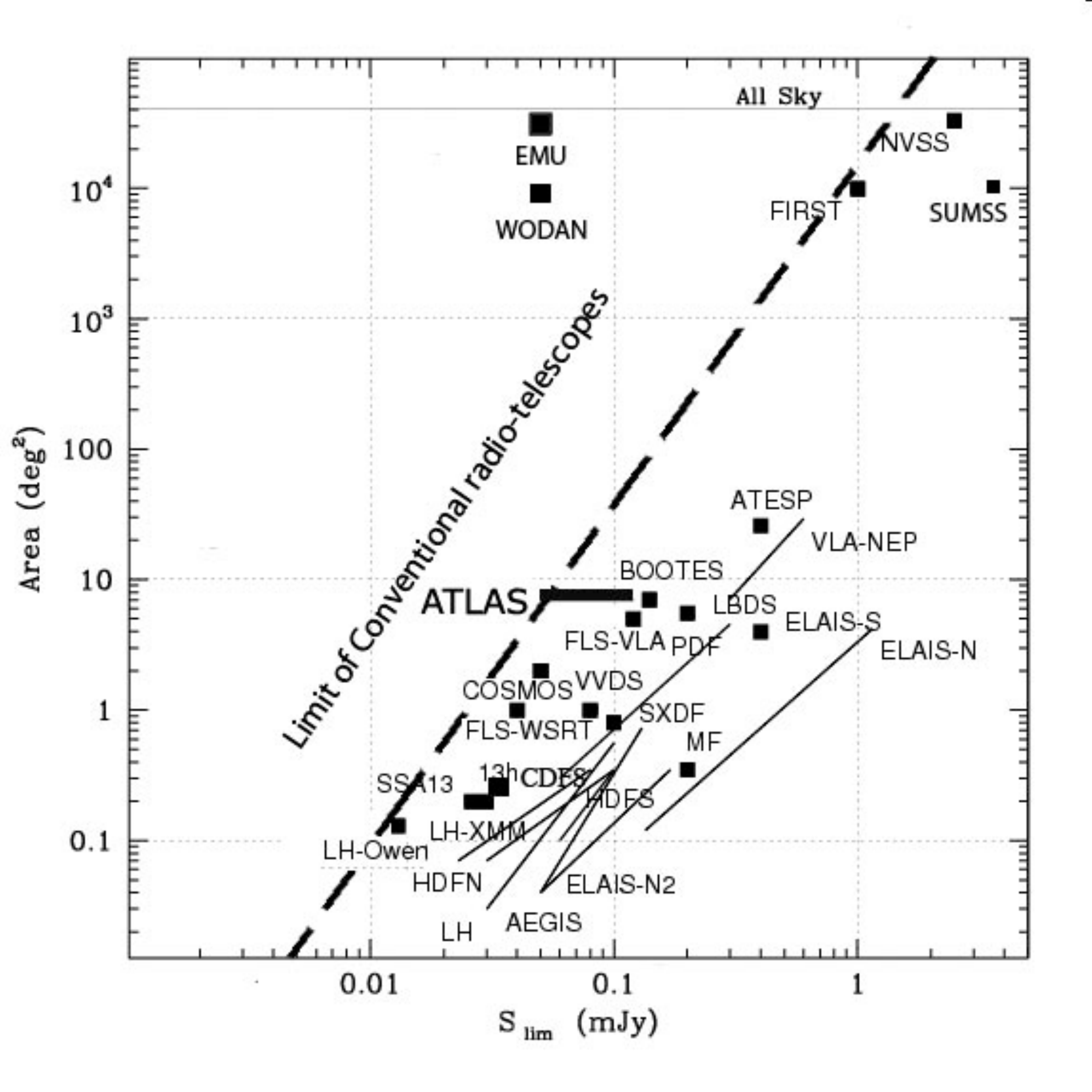}
\caption{Comparison of EMU with existing deep 20 cm radio surveys. Horizontal axis is 5-$\sigma$ sensitivity, and vertical axis shows the sky coverage. The largest existing radio survey is the wide but shallow NRAO VLA Sky Survey  (Condon et al., 1998) in the top right. The most sensitive radio survey is the deep but narrow Lockman Hole observation  (Owen et al., 2008) in the lower left. The squares at top centre represent the EMU survey, discussed here, and the complementary Northern WODAN survey. }
\label{fig1}
\end{center}
\end{figure}

The primary goal of EMU is to make a deep (10 \ujybm\ rms) radio continuum survey of the entire Southern Sky, extending as far North as $+30\deg$. EMU will cover roughly the same fraction (75\%) of the sky as NVSS  (Condon et al. 1998), but will be 45 times more sensitive, and will have an angular resolution (10 arcsec) 4.5 times better, as well as having higher sensitivity to extended structures. 
EMU is expected to generate a catalogue of about \ngal million galaxies, some 40 times greater than NVSS. Of these, about \nsf million are expected to be star-forming galaxies  at redshifts up to z$\sim$2, and the remainder AGNs up to z$\sim$5 (Norris et al., 2011).

In the Northern hemisphere, EMU will be complemented by the WODAN survey  (R\"ottgering et al., 2010) which has been proposed for the upgraded Westerbork telescope. WODAN will cover the northern 25\% of the sky (i.e. North of declination $+30\deg$) that is inaccessible to ASKAP, with a  small overlap for consistency and calibration checks. Together, EMU and WODAN will provide a full-sky 1.4\,GHz survey at $\sim$ 10--15 \arcsec\ resolution to an rms noise level of 10 \ujybm. 

Whilst previous large surveys such as NVSS were dominated by radio-loud active galactic nuclei (AGN), surveys at this sensitivity level are dominated by  star-forming galaxies (Seymour et al., 2008).
Thus, whereas most traditional radio-astronomical surveys had their greatest impact on the niche area of radio-loud AGN, EMU will be dominated by the same star-forming galaxies as are studied by optical and IR surveys, making it an important component of multi-wavelength studies of galactic evolution. Consequently, we also plan to cross-identify the EMU radio sources with sources in major optical/IR surveys as part of the EMU project.

We are fortunate in having access to earlier surveys such as  ATLAS  (Norris et al., 2006; Middelberg et al., 2008) and HDFS  (Norris et al., 2005; Huynh et al., 2005) with a sensitivity and resolution (and dynamic range challenges!) similar to those of EMU, but over a much smaller survey area. We are therefore using these  as a test-bed for the EMU Design Study,  and the prototype EMU source extraction and identification pipeline will be used for the final ATLAS data release in late 2011 (Banfield et al., 2011). 

\begin{figure}[h]
\begin{center}
\includegraphics[width=12cm, angle=0]{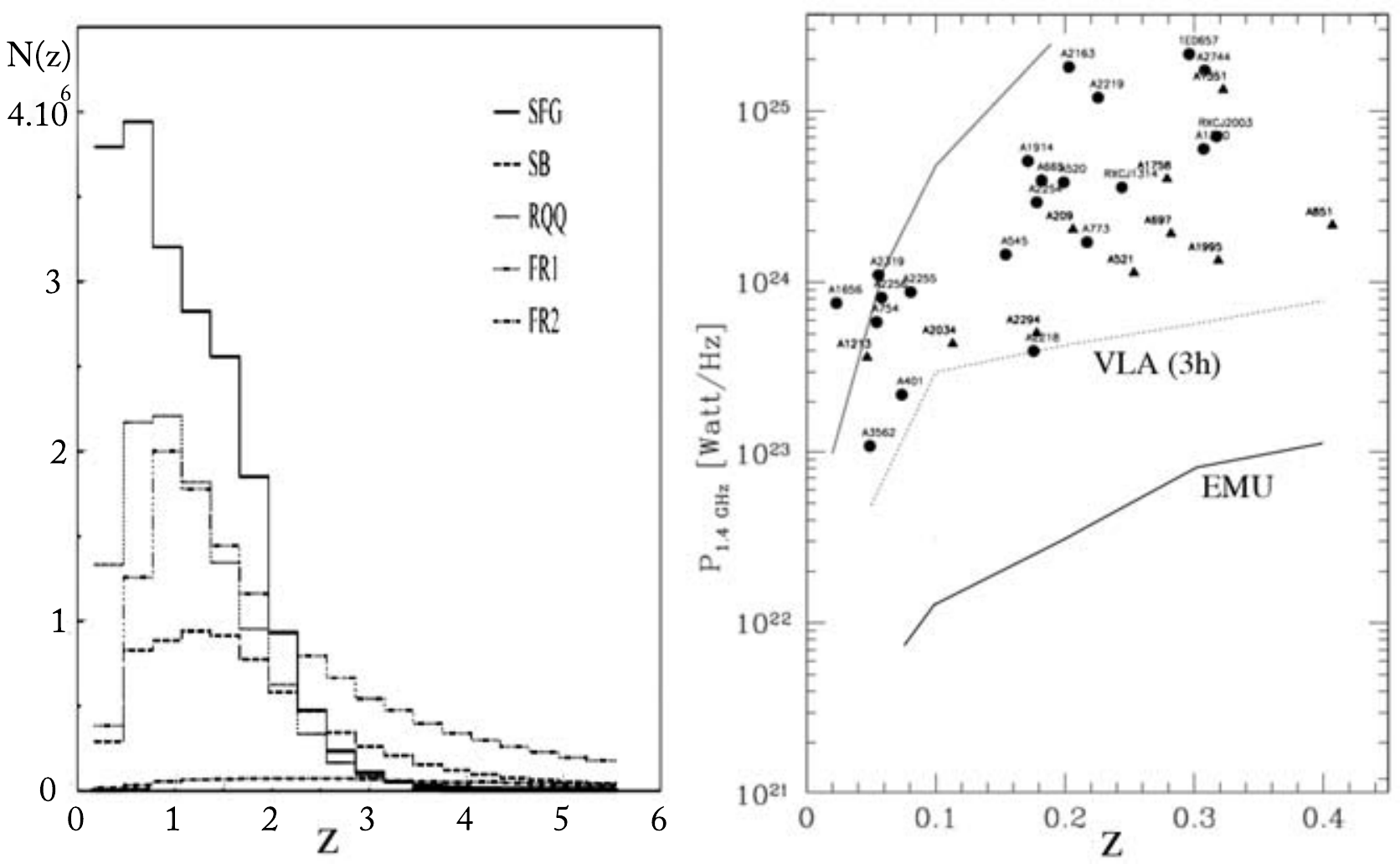}
\caption{{\em (Left)} Estimate of the redshift distribution of EMU sources (Norris et al., 2011). The four lines indicate star-forming galaxies (SFG), starburst galaxies (SB), radio-quiet quasars (RQQ), and radio-loud galaxies of Fanaroff-Riley types I and II. The mean redshift for EMU-detected AGNs is z = 1.88 for AGNs, and z = 1.1 for star-forming galaxies.
{\em (Right)} 1.4 GHz radio power of detected cluster haloes as a function of redshift showing the detection limits of previous cluster observations, adapted from Giovannini et al., (2009), and the calculated 5-$\sigma$ detection limit of EMU, assuming a halo with a  diameter of 1 Mpc, using the calculated ASKAP sensitivity on different scale sizes given by Norris et al. (2011). The upper line shows the limit corresponding to a scale size of 15 arcmin, which is approximately the largest size object that can be imaged with the VLA unless single-dish data is added to the interferometry data. At high redshifts, sensitivity may be limited by confusion, although we expect to overcome this by subtracting off compact sources.
 }
\label{nz}
\label{clusters}
\end{center}
\end{figure}

The key science goals for EMU are:
\begin{itemize}
\item To trace the evolution of star-forming galaxies from $z=2$ to the present day, using a wavelength unbiased by dust or molecular emission.
\item To trace the evolution of massive black holes throughout the history of the Universe, and understand their relationship to star formation.
\item To use the distribution of radio sources to explore the large-scale structure and cosmological parameters of the Universe, and to test fundamental physics.
\item To explore an uncharted region of observational parameter space, with a high likelihood of finding new classes of object.
\item To use radio sources to trace clusters and large-scale structure, and explore the astrophysics of dark matter haloes.\item To create the most sensitive wide-field atlas of Galactic continuum emission yet made in the Southern Hemisphere, addressing areas such as star formation, supernovae, and Galactic structure.
\end{itemize}

The remainder of this paper focuses on just one of these science goals: the study of clusters of galaxies.

\section{Clusters of Galaxies}
The study of clusters of galaxies has changed significantly in recent years with the realisation that, rather than being isolated entities, they represent the intersections of filaments and sheets in the large-scale structure of the Universe, as represented by the Millennium simulation (Springel et al., 2005). Cluster studies therefore aim not only to understand the physics of clusters  themselves, but also to trace the evolution of structure in the Universe. 

Clusters are typically found either through X-ray searches (Rosati et al. 1998; Romer et al. 2001; Pierre et al. 2003) or by using optical colour as a surrogate for redshift, enabling searching for clusters in colour-position space (Gladders \& Yee 2005; Wilson et al. 2008; Kodama et al. 2007). As a result,  tens of thousands of clusters are currently known, but only a few at z$ > $ 1 (Wilson et al. 2008, Kodama et al. 2007), with the highest redshift at z = 2.07 (Gobat et al., 2011).

At radio wavelengths, clusters of galaxies are characterised by three different powering mechanisms  (Kempner et al., 2004) in addition to the radio emission from their constituent galaxies:
\begin{itemize}
\item haloes at the centres of clusters, 
\item relics (representing shocks from cluster-cluster collisions) at the periphery, and 
\item tailed radio galaxies, which are an important tracer and barometer of the intra-cluster medium.  
\end{itemize} 
Not only are these three types of radio source  important as tracers of clusters, but all three are diagnostics of the physics of clusters, particularly when combined with X-Ray data. 
However, the number of currently detected cluster radio sources is limited by the present telescope sensitivities (see Fig \ref{clusters}). 
EMU will push beyond the present limits to detect diffuse sources with a range of powers over a larger redshift range, greatly improving our understanding of these sources.

\subsection{Haloes}
Diffuse synchrotron radio haloes are found in clusters and groups of galaxies, indicating strong magnetic fields and relativistic particles, presumably accelerated by the released potential energy of cluster formation. Only  a few tens of radio haloes are known, and are typically discovered by making deep radio surveys of X-ray-detected haloes (Venturi et al. 2008).
The ATLBS survey  (Subrahmanyan et al., 2010), which has surveyed 8.4\sqdeg\  to an rms sensitivity of 80 \ujybm\  on a scale size of 50\arcsec\  at 1.4 GHz, has detected tens of diffuse sources, of which about 20 have been tentatively identified as cluster and group haloes  (Saripalli et al., 2011). If these numbers are confirmed, then EMU, with significantly greater sensitivity to extended structures than the LBS survey, should detect $\sim$ 60000 cluster and groups haloes, which dramatically increases the number of known clusters.
Brunetti et al. (2009), Cassano et al. (2010), and Schuecker et al. (2001) have suggested that radio haloes in the centres of clusters are distributed bimodally as a function of X-ray luminosity, with haloes generally found only in those clusters which have recently undergone a merger, resulting in a disturbed appearance at X-ray wavelengths. As well as discovering new clusters, EMU will provide a uniform radio sample of emission across all clusters   which will allow us to test 
whether very faint radio haloes occur in all galaxy clusters.

\subsection{Relics}
On the periphery of clusters, elongated radio ``relics'' are found, typically oriented perpendicular to the radius vector from the cluster centre. These are interpreted as  shock structures (Brown et al., 2011; Markevitch et al., 2010; van Weeren et al., 2010) and provide important diagnostics for the dynamics of accretion and mergers of clusters  (Barrena et al., 2009). 
Only 44 radio relics are currently known  (Giovannini et al., 2011), and few have been discovered in current radio surveys because of the relatively poor sensitivity of most surveys to low-surface-brightness structures. One has been discovered in the seven square degrees of ATLAS (Middelberg et al., 2008; Mao et al. 2010), at z $\sim$ 0.2. EMU will have greater sensitivity to such low-surface-brightness structures than ATLAS, and so in the 30000 \sqdeg\ of EMU we expect to detect $>$ 4000, although this number is clearly very uncertain.

\subsection{Tailed radio galaxies}
Tailed radio galaxies are found in large clusters, and are believed to represent radio-loud AGN in which the jets are distorted by the intra-cluster medium  (Mao et al., 2010). Mao et al. (2011b) have found 6 tailed galaxies in ATLAS, from which they estimate between $26\times 10^3$ and $2\times 10^5$ tailed galaxies will be detected by EMU, depending on their luminosity function and density evolution. Deghan et al. (2011), using high-resolution images of one of the ATLAS fields, find 12 tailed galaxies in 4 \sqdeg, implying $\sim 10^5$ tailed galaxies in EMU. Importantly, such galaxies can be detected out to high redshifts (Wing et al., 2011; Mao et al., 2010), providing a powerful diagnostic for finding clusters. 

\section{Redshifts and Polarisation}
To interpret data from EMU, redshifts are invaluable. However, no existing or planned redshift survey can cover more than a tiny fraction of EMU's \ngal million sources. For nearby galaxies, HI redshifts will be available from WALLABY, which will provide $\sim 5 \times 10^5$ redshifts, and smaller numbers will be provided by other redshift surveys such as SDSS (York et al., 2000) and GAMA (Driver et al., 2009). The remaining $\sim$ 99\% of EMU galaxies will not have spectroscopic redshifts.

Photometric redshifts, in which SEDs of template galaxies are fitted to the measured multi-band photometry of target galaxies, are often used as a surrogate for spectroscopic redshifts. 
While the relatively sparse photometry available for most EMU sources can not generate accurate photometric redshifts, it will enable \emph{estimates} to be made for about 30\% of EMU galaxies. Even a non-detection can carry useful information, and radio data themselves can add significantly to the choice of SED template, and hence to a probabilistic estimate of redshift. For example,  high-redshift radio galaxies can be identified from their strong radio emission coupled with a K-band non-detection (Willott et al., 2003). The radio data alone can also weight the probability of a particular redshift range. For example, a steep radio spectral index increases the probability of a high redshift (Breuck et al., 2002), while the angular size of a particular galaxy class can be loosely correlated with redshift (Wardle \& Miley, 1974). 

Fortunately, for many purposes, such as cosmological tests (Raccanelli et al., 2011), approximate redshifts are sufficient. In such ``statistical redshifts'', only a fraction of objects will have an approximately correct redshift, the remaining incorrect redshifts merely generating noise which can be cancelled in a statistical study of a sufficient number of objects. It is important, for this purpose, that reliability and completeness are carefully calibrated in a small well-studied area with deep spectroscopic redshifts. 

Polarisation data  can also be used to make statistical statements about redshifts. POSSUM  (Gaensler et al., 2010) is an ASKAP project that will run commensally with EMU, generating a catalogue of polarised fluxes and Faraday rotation measures (or upper limits) for all sources detected by EMU. POSSUM data will help us to distinguish (e.g.) the largely polarised tailed galaxies and relics from the largely unpolarised halo emission.
 Compact sources that are strongly polarised are nearly always AGN (Hales et al., 2011), and so have a mean $z \sim 1.8$, while unpolarised sources are mainly star-forming galaxies with a mean $z \sim 0.8$. Consequently, cosmological tests may be made by treating unpolarised sources as a low-redshift screen in front of background high-redshift polarised sources.

\section{Discussion and Conclusion}
EMU will provide an unbiased radio survey for cluster haloes, relics, and tailed galaxies. It will be important to compare  the properties
of these radio-selected clusters to those of the X-ray selected
population from surveys such as the eROSITA all-sky X-ray survey (Predehl et al., 2010), and the  Sunyaev-ZelÕdovich surveys made with the South Pole Telescope (Williamson et al., 2011), Atacama Cosmology Telescope (Marriage et al., 2010) and Planck (Planck Collaboration, 2011) surveys, which are biased towards the detection of high-mass clusters at high redshifts.

Quite apart from the value of an unbiased radio survey  of known clusters, EMU will detect at least $3 \times 10^4$, and possibly hundreds of thousands, of new clusters, roughly doubling the number of known clusters. This is particularly important at redshifts z$>$0.5 where current constraints on large-scale structure are weaker, and traditional detection techniques like X-ray surveys and use of the Red Cluster Sequence (Gladders \& Yee 2005) become less effective, while a few clusters have already been detected using radio techniques up to high redshifts (Blanton et al. 2003, Wing et al. 2011).

In principle, radio techniques can be used to detect clusters even beyond z=1. While the highest-redshift tailed galaxy currently known (z = 0.96; Blanton et al. 2003) is detectable by EMU up to z $\sim$ 3, high-resolution follow-up will be needed to confirm an initial EMU candidate list. But extrapolation beyond z=1 is uncertain: we do not know the luminosity function of these galaxies, nor their evolution. Furthermore, inverse Compton cooling of electrons by the cosmic microwave background is expected to quench their synchrotron radio emission at $z \gg 1$, although the same mechanism would also be expected to distort the far-IR-radio correlation at high redshift, which is not observed (Mao et al. 2011a).

In summary, EMU will deliver a next-generation radio survey covering three quarters of the sky to sensitivity levels that are currently reached only by intensive small-area surveys. It will remain the largest radio survey until the 
SKA, with significant impact on many areas of astronomy. EMU will be especially productive for cluster research, and may discover more clusters than any other technique. The deep unbiased survey will also be invaluable for comparison with cluster properties at X-ray and other wavelengths.
 
To ensure that the science goals are achieved, it is necessary to work carefully through each of them in detail before the survey starts in 2013, ensuring that the survey strategy is optimised to generate the science. To achieve this,  we welcome interactions and collaborations with observers and theoreticians, and with other surveys, to increase our common scientific productivity.  Further information on EMU can be found on http://askap.org/emu

\section*{Acknowledgments}
The EMU team consists of over 180 members from 14 countries, all of whom are listed on
\small{http://askap.pbworks.com/TeamMembers}. This paper draws on the expertise and ideas of all of them. I particularly thank the coauthors of Norris et al. (2011) for their permission to incorporate the ideas expressed in that paper.

\label{lastpage}
\end{document}